\documentclass[12pt]{iopart}

\usepackage{graphicx}
\usepackage{hyperref}
\usepackage{color}
\usepackage{cancel}
\usepackage{soul}

\expandafter\let\csname equation*\endcsname\relax
\expandafter\let\csname endequation*\endcsname\relax
\usepackage{amsmath}

\begin{document}

\title{Transverse momentum spectra of hadrons in high energy pp and heavy ion collisions}

\author{Kapil Saraswat$^{1, \ast}$, Prashant Shukla$^{2,3, \dagger}$ and Venktesh Singh$^{1}, \ddagger$}
\address{$^{1}$Department of Physics, Institute of Science, Banaras Hindu University, Varanasi 221005, India.}
\address{$^{2}$Nuclear Physics Division, Bhabha Atomic Research Center, Mumbai 400085, India.}
\address{$^{3}$Homi Bhabha National Institute, Anushakti Nagar, Mumbai 400094, India.}
\ead{$^{\ast}$ kapilsaraswatbhu@gmail.com}
\ead{$^{\dagger}$ pshuklabarc@gmail.com}
\ead{$^{\ddagger}$ venkaz@yahoo.com}

\begin{abstract}
 We present a study of transverse momentum ($p_{\rm{T}}$) spectra of unidentified 
 charged particles in pp collisions at RHIC and LHC energies from
 $\sqrt{s} = 62.4$~GeV to 13~TeV using Tsallis/Hagedorn function. The power law of 
Tsallis/Hagedorn form gives very good description of the hadron spectra in 
$p_{T}$ range from 0.2 to 300 GeV/$c$. The power index $n$ of the $p_{\rm{T}}$ 
distributions is found to follow a function of the type $a+b/\sqrt {s}$ with 
asymptotic value $a = 6.8$. The parameter $T$ governing the soft bulk 
contribution to the spectra remains almost same over wide range of collision
energies. We also provide a Tsallis/Hagedorn fit to the $p_{\rm{T}}$ spectra of 
hadrons in pPb and different centralities of PbPb collisions at
$\sqrt{s_{\rm NN}} = 5.02$~TeV. The data/fit shows deviations from the Tsallis distribution which 
become more pronounced as the system size increases. We suggest simple 
modifications in the Tsallis/Hagedorn power law function and show that 
the above deviations can be attributed to the transverse flow in low 
$p_{\rm{T}}$ region and to the in-medium energy loss in high $p_{\rm{T}}$ region.   
\end{abstract}

%
\submitto{Journal of Physics Communications}
\maketitle

\section{Introduction}

 The light hadrons are the most abundant particles produced in the pp and heavy
ion collisions. The transverse momentum ($p_{\rm{T}}$) spectra of hadrons can be used
to infer the particle production mechanism in pp collisions. In heavy ion
collisions, additional final state effects such as collective flow
\cite{Hirano:2003pw, Fries:2008hs}, recombination \cite{Fries:2008hs, Fries:2004ej}
and jet-quenching \cite{Wang:2003aw} in different $p_{\rm{T}}$ ranges are superimposed
over the hadron spectra. The hadron $p_{\rm{T}}$ spectra in pp collisions are successfully
described by Tsallis distribution ~\cite{Tsallis:1987eu, Biro:2008hz} in terms
of only two parameters, the Tsallis parameter $T$ and the parameter $q$ which governs
the degree of non-thermalization. It is well known 
\cite{Khandai:2013gva, Wong:2012zr, Wong:2013sca} that the functional form of the
Tsallis distribution which describes near-thermal systems is essentially the same
as the power law function by Hagedorn which is applicable to QCD hard scatterings
~\cite{Hagedorn:1983wk, Blankenbecler:1974tm}.  There are numerous studies which
show that the Tsallis/Hagedorn distribution gives an excellent description of
$p_{\rm{T}}$ spectra of all identified hadrons measured in pp collisions at SPS, RHIC
and LHC energies \cite{Khandai:2013gva, Adare:2010fe, Sett:2014csa, Zheng:2015tua}.
 The work in Ref.~\cite{Zheng:2015mhz} makes a comparative study of various forms of
Tsallis distributions available in the literature by fitting the hadron $p_{\rm{T}}$
spectra measured at RHIC and LHC. The recent papers
\cite{Parvan:2016rln, Parvan:2016mbv, Azmi:2014dwa}
present a study of the Tsallis parameters for $p_{\rm{T}}$ distributions of pions produced
in pp collisions as a function of $\sqrt{s}$ ranging between 6.3 GeV and 7 TeV.
The CMS papers \cite{Chatrchyan:2012qb, Sirunyan:2017zmn} present studies
on measured identified particle spectra using Tsallis distribution
at $\sqrt{s}$ = 0.9, 2.76, 7 and 13 TeV.
There are many attempts to use the Tsallis distribution in heavy ion collisions after
taking into account the transverse collective flow which is a final state effect
~\cite{Tang:2008ud, Khandai:2013fwa, Sett:2015lja}. 
The work in Ref.~\cite{Saraswat:2017gqt} studies the $p_{\rm{T}}$ spectra of the strange hadrons
production in pp collision at $\sqrt{s}=$ 7 TeV, pPb collision at $\sqrt{s_{\rm{NN}}}=$ 
5.02 TeV and PbPb collision at $\sqrt{s_{\rm{NN}}}=$ 2.76 TeV using the Tsallis distribution
which includes the transverse flow.

 The Tsallis distribution is applied to unidentified light charged hadron spectra
measured in pp collisions at LHC over a wide $p_{\rm{T}}$ range upto 200 GeV/$c$
\cite{Cleymans:2015lxa}. The work in Ref.~\cite{Wong:2015mba} studies the $p_{\rm{T}}$
spectra of both jets and hadrons in pp collisions. They find that the power index
$n$ is 4-5 for jet production and  6-10 for hadron production. The work in Refs.
\cite{Wilk:2014sza, Wilk:2014bia} uses Tsallis power law to fit the hadron spectra
in wide $p_{\rm{T}}$ range measured in pp collisions. Looking more closely at the data/fit
they suggest that deviations of the data from power law fit follows a log-periodic
oscillation which could imply a complex exponent  of the power law. It was further
suggested~\cite{Rybczynski:2014ura, Wilk:2015pva} that the oscillations in data/fits
in PbPb collisions at $\sqrt{s_{\rm{NN}}}$ = 2.76 TeV are similar to those in pp data in
the same range of transverse momenta.

  In this work, we study transverse momentum ($p_{\rm{T}}$) spectra of unidentified charged
particles in pp collisions at  $\sqrt{s}$ = 0.0624, 0.2, 0.9, 2.36, 2.76, 5.02, 7 and
13 TeV using Tsallis/Hagedorn function.
The statistical and systematic errors are added in quadrature and are used in the fits.
The parameters of such fits are studied as a function of beam energies.
We  also study the spectra of identified charged particles in pp collisions
albeit at smaller $p_{\rm{T}}$. Since the aim of the work is to obtain a
function describing the hadron spectra in wide $p_{\rm{T}}$ range,  we choose the
unidentified particles for which the measurements are available at very high
$p_{\rm{T}}$ upto 200~GeV/$c$. 
 We also make a Tsallis/Hagedorn fit to the $p_{\rm{T}}$ spectra of hadrons in pPb and different
centralities of PbPb collisions at $\sqrt{s_{\rm{NN}}}$ = 5.02 TeV. We suggest simple
modifications in the Tsallis/Hagedorn power law function to include transverse flow
and in-medium energy loss in the hadronic spectra.

\section{Tsallis/Hagedorn distribution function and the modification}

The transverse mass ($m_{\rm{T}} =\sqrt{p_{\rm{T}}^2+m^2}$) distribution of particles produced in
hadronic collisions can be described by the Hagedorn function which is a QCD-inspired
summed power law \cite{Hagedorn:1983wk} given as
\begin{eqnarray}
E~\frac{d^{3}N}{dp^{3}} = A~ \Bigg(1 + \frac{m_{\rm{T}}}{p_{0}}\Bigg)^{-n}~.
\label{Hag}
\end{eqnarray}
This function describes both the bulk spectra in the low $m_{\rm{T}}$ region and the particles
produced in QCD hard scatterings reflected in the high $p_{\rm{T}}$ region. Let us compare this
function with the Tsallis distribution \cite{Tsallis:1987eu, Biro:2008hz} of thermodynamic
origin given by
\begin{eqnarray}
E \frac{d^{3}N}{dp^{3}} = C_n ~ m_{\rm{T}} ~ \Bigg(1 + (q-1) \frac{m_{\rm{T}}}{T} \Bigg)^{-1/(q - 1)}~.
\label{Tsallis}
\end{eqnarray}
 The Tsallis distribution describes near-thermal systems in terms of Tsallis parameter $T$ and
the parameter $q$ which measures degree of non-thermalization \cite{Wilk:1999dr}. The functions
in Eq.~\ref{Hag} and in Eq.~\ref{Tsallis} have similar mathematical forms with $n = 1/(q - 1)$ and
$p_0 = n\,T$. Larger values of $n$ correspond to smaller values of $q$. Both $n$ and $q$ have been
interchangeably used in Tsallis distribution
\cite{Biro:2008hz, Adare:2010fe, Cleymans:2012ya, Adare:2011vy, Abelev:2006cs}.
Phenomenological studies suggest that, for quark-quark point scattering, $n\sim4$
\cite{Blankenbecler:1975ct, Brodsky:2005fza}, which grows larger if multiple scattering centers
are involved.
 The study in Ref.~\cite{Zheng:2015mhz} suggests that both the forms given in 
Eq.~\ref{Hag} and in Eq.~\ref{Tsallis} give equally good fit to the hadron spectra
in pp collisions. We use Eq.~\ref{Tsallis} in case of pp collisions.

 Tsallis/Hagedorn function is able to describe $p_{\rm{T}}$ spectra in pp collisions practically
at all generations of proton colliders. There have been many attempts to use the Tsallis
distribution in heavy ion collisions as well by including the transverse collective flow
~\cite{Tang:2008ud, Khandai:2013fwa, Sett:2015lja}. In addition, in heavy ion collisions,
particle spectra at high $p_{\rm{T}}$ are known to be modified due to in-medium energy loss.
The Tsallis/Hagedorn distribution can be modified by including these final state effects in
different $p_{\rm{T}}$ regions as follows:

\begin{subequations} \label{modified_new_func_tsallis_distribution_function}
\begin{align} 
E \frac{d^{3}N}{dp^{3}} &= A_{1} \Bigg[\exp\left(-\frac{\beta  p_{\rm{T}}}{p_{1}}\right)
   + \frac{m_{\rm{T}}}{p_{1}}\Bigg]^{-n_{1}} ~ :~  p_{\rm{T}} < p_{\rm{T_{th}}} ~.
\label{new_func_tsallis_distribution_function} \\
E \frac{d^{3}N}{dp^{3}}  &= A_{2}~ \Bigg[\frac{B}{p_{2}}~\Bigg(\frac{p_{\rm{T}}}{q_{0}}\Bigg)^{\alpha}
                              + \frac{m_{\rm{T}}}{p_{2}}\Bigg]^{-n_{2}}~ :~  p_{\rm{T}} > p_{\rm{T_{th}}} ~. 
 \label{new_func_tsallis_distribution_function_second}
\end{align} 
\end{subequations}
The first function (Eq.~\ref{new_func_tsallis_distribution_function}) is shown to govern the thermal
and collective part of the hadron spectrum with the temperature $T=p_{1}/n_{1}$ and the average transverse
flow velocity $\beta$~\cite{Khandai:2013fwa}.

The second function (Eq.~\ref{new_func_tsallis_distribution_function_second})
is obtained after shifting the distribution in Eq.\ref{Hag} by energy  loss
$\Delta m_{\rm{T}}$ in the medium as 
\begin{equation}
  E \frac{d^{3}N}{dp^{3}} = A_2~ \Bigg[1 + \frac{m_{\rm{T}} +\Delta m_{\rm{T}}}{p_{2}}\Bigg]^{-n_{2}}~.
\label{tsallis_with_energyloss}
\end{equation}
The energy loss $\Delta m_{\rm{T}}$ is proportional to $p_{\rm{T}}$ at low $p_T$ 
and in general can be parameterized similar to
the work in Ref.~\cite{Spousta:2016agr} as
\begin{equation}
\Delta m_{\rm{T}} = B ~ \Big(\frac{p_{\rm{T}}}{q_{0}}\Big)^\alpha~. 
\label{spousta_energy_loss}
\end{equation}
 Here, the parameter $\alpha$ quantifies different energy
loss regimes for light quarks in the medium~\cite{Baier:2000mf, De:2011fe}.
 The parameter $B$ is proportional to the medium size and 
$q_{0}$ is an arbitrary scale set as 1~GeV. 
 Using Eq.~\ref{spousta_energy_loss} in Eq.~\ref{tsallis_with_energyloss} and
 ignoring 1 we get Eq.~\ref{new_func_tsallis_distribution_function_second}
 applicable for high  $p_T$.
 In our study, we find that this function describes the particle spectra
at $p_{\rm T_{th}}$ above 7~GeV/$c$. Fits to the data would constrain the
value of $B/p_2$ and thus $p_2$ is not an independent parameter. 
The empirical parton energy loss in nuclear collisions at RHIC
  energies is found to be proportional to $p_{\rm{T}}$ \cite{Wang:2008se}.

\section{Results and discussions}

\begin{figure}[htp] 
\center
\includegraphics[width=0.65\linewidth]{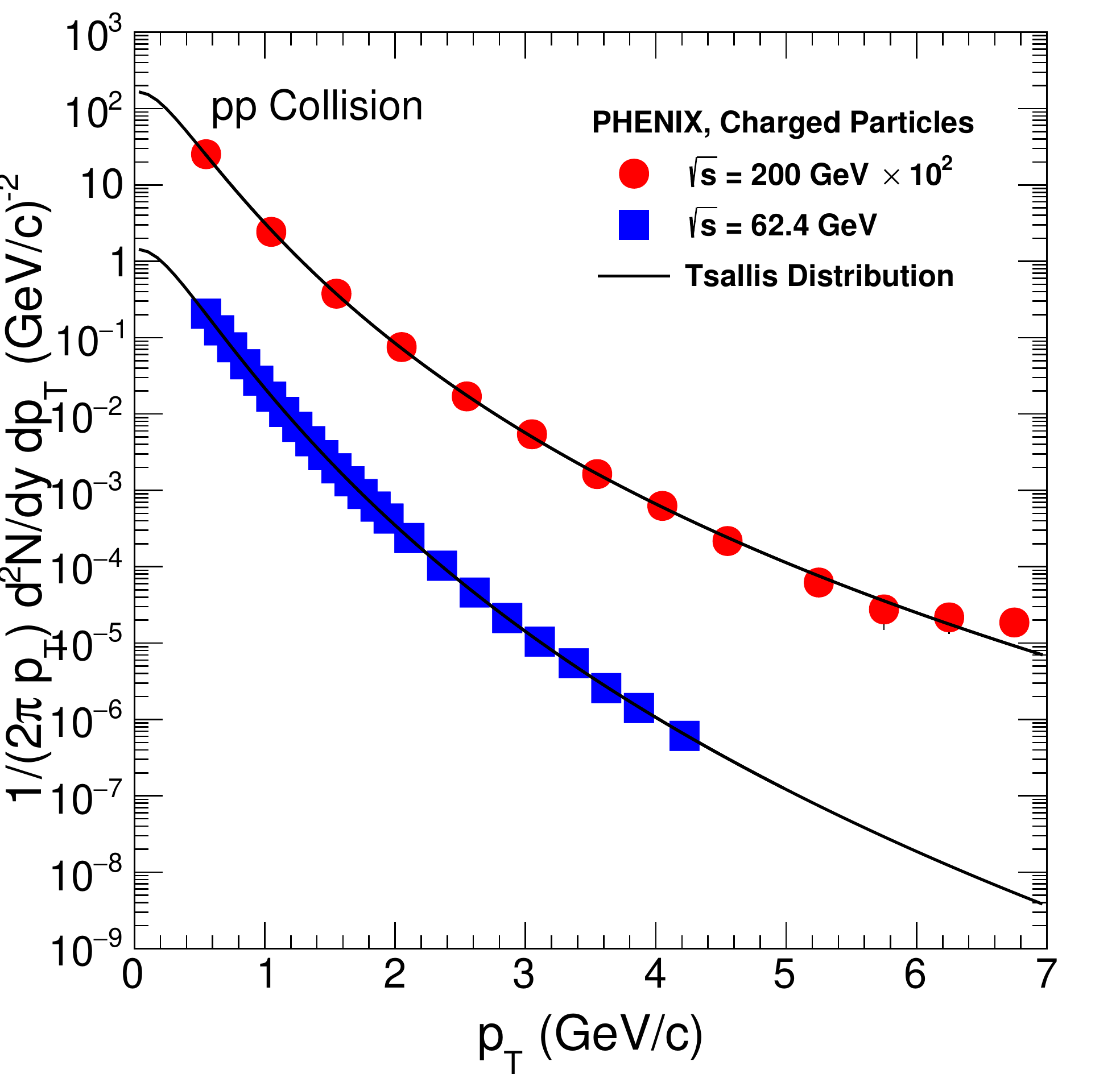}
\caption{The invariant yields of the charged particles as a function of transverse 
momentum $p_{\rm{T}}$ for pp collision at $\sqrt{s}$= 62.4 and 200 GeV measured by the 
PHENIX experiment \cite{Adare:2012nq, Adler:2005in}. The solid curves are the fitted 
Tsallis distributions.} 
\label{Figure1_pp_collision_phenix}  
\end{figure}

Figure \ref{Figure1_pp_collision_phenix} shows the invariant yields of the charged
particles as a function of $p_{\rm{T}}$ for pp collisions at $\sqrt{s}$ = 62.4 and 200 GeV
measured by the PHENIX experiment \cite{Adare:2012nq, Adler:2005in}. The solid curves
are the Tsallis distributions fitted to the spectra. The Tsallis distribution function
gives good description of the data for both the collision energies which can be inferred
from the values  of $\chi^{2}/\rm{NDF}$ given in the Table~\ref{pp_collision_chi2_NDF}.

\begin{figure}[htp] 
\center
\includegraphics[width=0.65\linewidth]{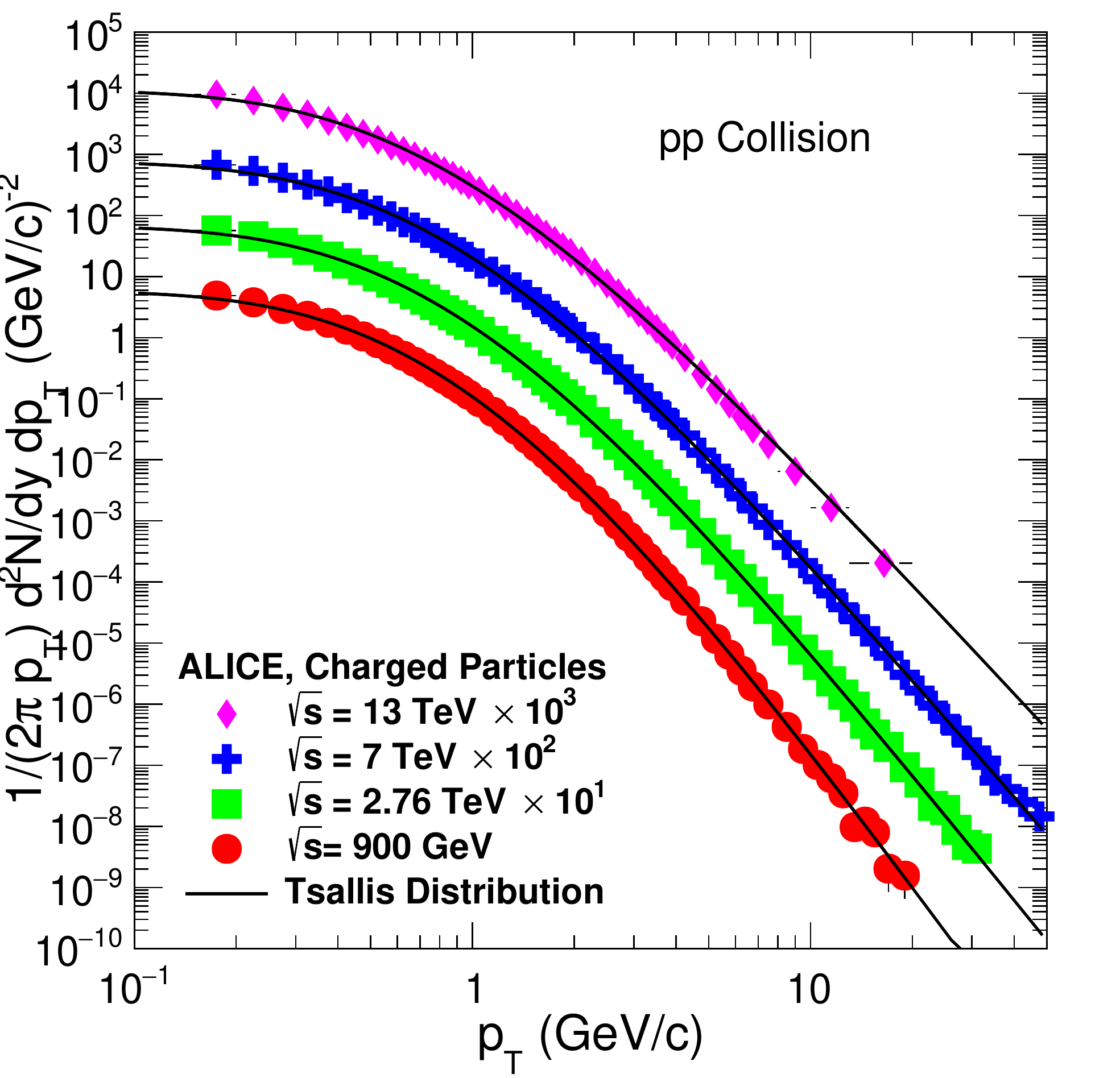}
\caption{The invariant yields of the charged particles as a function of transverse 
momentum $p_{\rm{T}}$ for pp collision at $\sqrt{s}$= 0.9, 2.76, 7 and 13 TeV measured 
by the ALICE experiment \cite{Abelev:2013ala, Adam:2015pza}. The solid curves are 
the fitted Tsallis distribution functions.} 
\label{Figure2_pp_collision_alice} 
\end{figure}

Figure \ref{Figure2_pp_collision_alice} shows the invariant yields of the charged
particles as a function of $p_{\rm{T}}$ for pp collisions at $\sqrt{s}$ = 0.9, 2.76, 7 and
13 TeV measured by the ALICE experiment \cite{Abelev:2013ala, Adam:2015pza}. The solid
curves are the Tsallis distributions. The Tsallis distribution function gives good
description of the data for collision energies which can be inferred from the values
of $\chi^{2}/\rm{NDF}$ given in the Table~\ref{pp_collision_chi2_NDF}.

\begin{figure}[htp] 
\center
\includegraphics[width=0.65\linewidth]{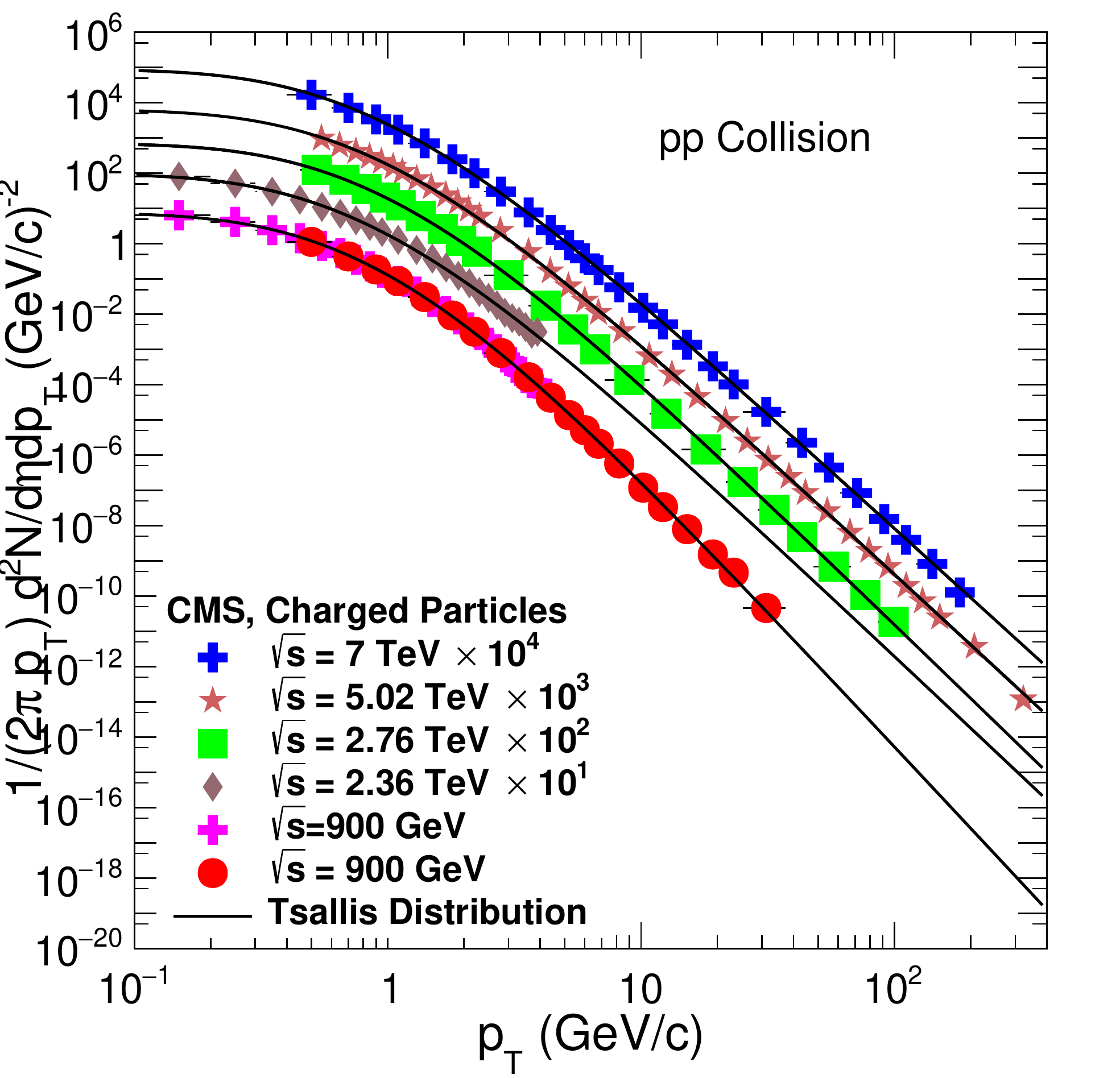}
\caption{The invariant yields of the charged particles as a function of transverse momentum 
$p_{\rm{T}}$ for pp collision at $\sqrt{s}$= 0.9, 2.36, 2.76, 5.02 and 7 TeV measured by the 
CMS experiment \cite{Khachatryan:2010xs, Khachatryan:2016odn, Chatrchyan:2011av}. The solid 
curves are the fitted Tsallis distribution functions.} 
\label{Figure3_pp_collision_cms} 
\end{figure}

Figure \ref{Figure3_pp_collision_cms} shows the invariant yields of the charged 
particles as a function of $p_{\rm{T}}$ for pp collisions at $\sqrt{s}$ = 0.9, 2.36,
2.76, 5.02 and 7 TeV measured by the CMS experiment 
\cite{CMS:2012aa, Khachatryan:2010xs, Khachatryan:2016odn, Chatrchyan:2011av}.
The solid curves are the Tsallis distributions fitted to the spectra. The
Tsallis distribution function gives good description of the data for all collision
energies which can be inferred from the values of $\chi^{2}/\rm{NDF}$ gives in the
Table~\ref{pp_collision_chi2_NDF}.

\begin{figure}[htp] 
\center
\includegraphics[width=0.65\linewidth]{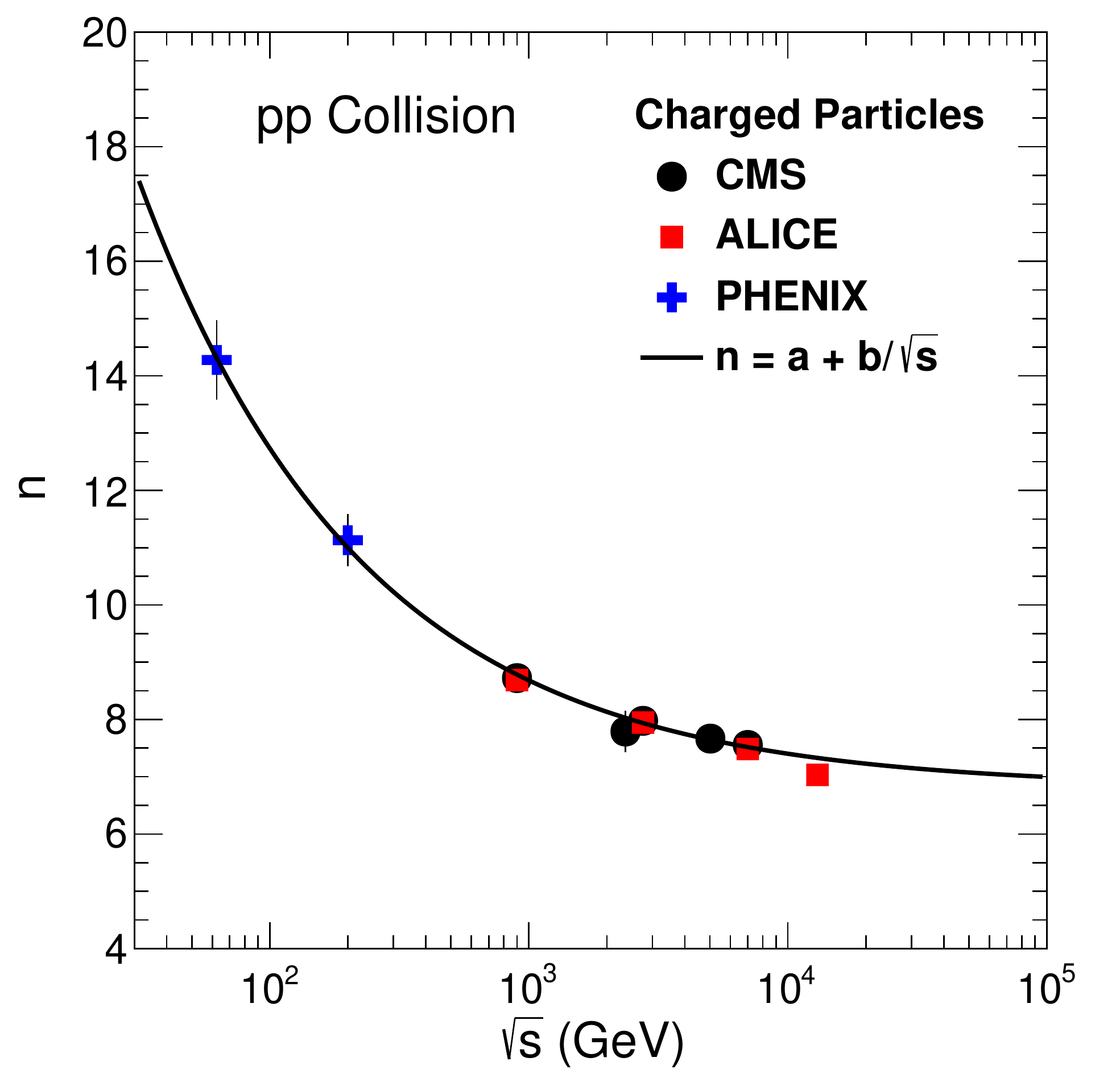}
\caption{The Tsallis parameter $n$ for the charged particle as a function of the centre of mass energy 
$\sqrt{s}$ of pp collision. The solid curve represents the function ($a + b/\sqrt{s}$).}
\label{Figure4_pp_collision_nn} 
\end{figure}

 Figure \ref{Figure4_pp_collision_nn} shows the Tsallis parameter $n$ for the charged
particles as a function of the pp collision energy $\sqrt{s}$. The value of $n$
decreases as we move from RHIC to LHC energies. The decreasing value of $n$ shows that
number of quarks participating for a produced particle
are reduced for higher energy collisions.
 The parameter $n$ can be parametrized by a function of the type
\begin{equation}
n (\sqrt{s}) = a + \frac{b}{\sqrt{s}}~.
\label{nn_parameterized_equation} 
\end{equation}
Here $a$ = 6.81 $\pm$ 0.06 and $b$ = 59.24  $\pm$ 3.53 GeV with $\chi^2$/NDF = 0.78.
The QCD point scattering for pions production gives $n=4$. Earlier studies have suggested
that the value of $n$ is larger for baryons as compared to that for mesons. Since the bulk
of the particles produced in the pp collisions are predominantly pions, we can consider
the unidentified particle spectra as that of pions.

\begin{figure}[htp] 
\center
\includegraphics[width=0.65\linewidth]{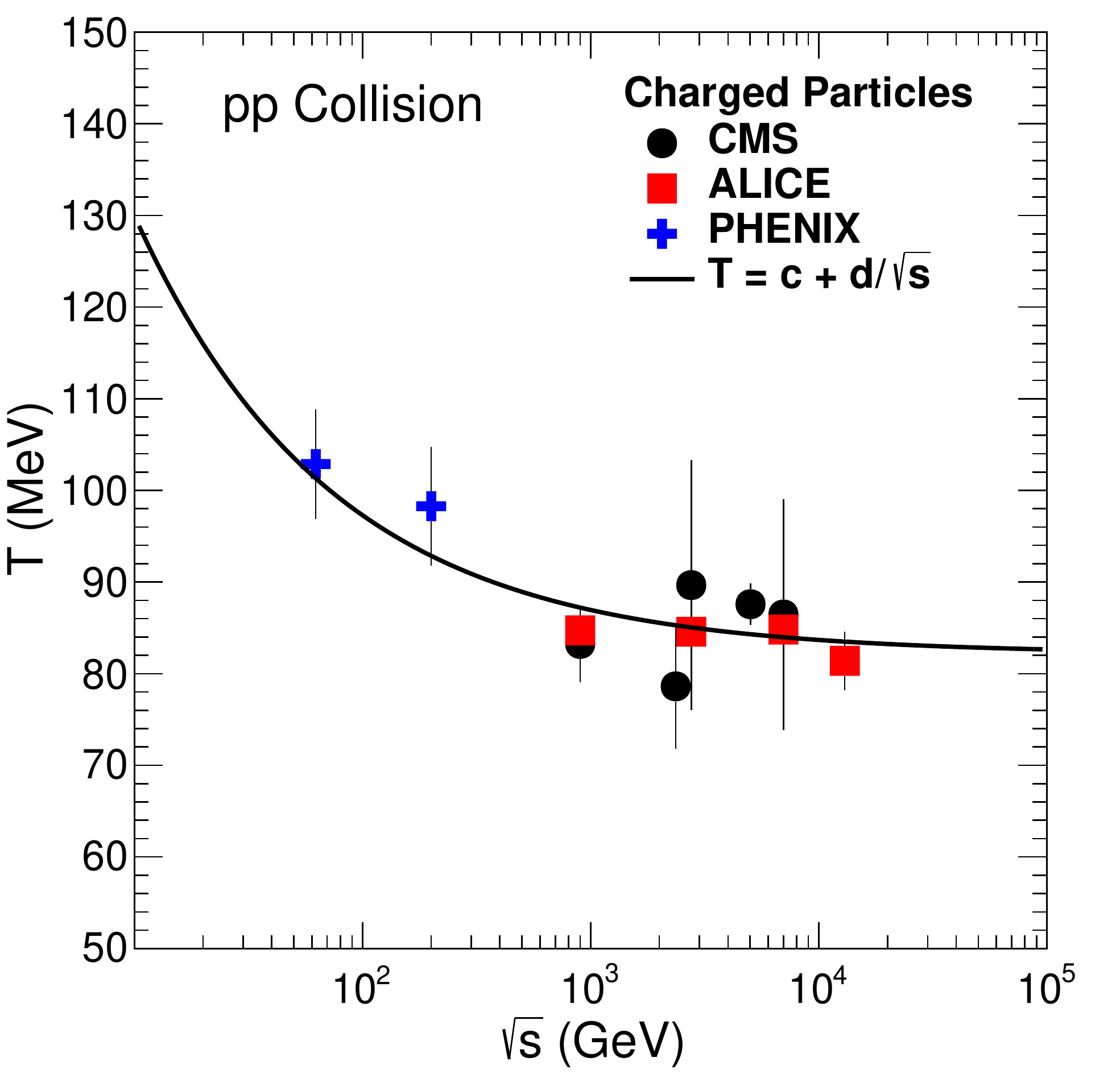}
\caption{The Tsallis temperature parameter $T$ for the charged particle as a function of the centre of mass energy 
$\sqrt{s}$ of pp collision.}
\label{Figure5_pp_collision_TT} 
\end{figure}

  Figure \ref{Figure5_pp_collision_TT}  shows the Tsallis temperature parameter $T$ for the charged
particles as a function of $\sqrt{s}$ in pp collision.
The parameter $T$ can be parametrized by a function of the type
\begin{equation}
T (\sqrt{s}) = c + \frac{d}{\sqrt{s}}~.
\label{TT_parameterized_equation} 
\end{equation}
Here $c$ = 0.082 $\pm$ 0.002 GeV and $d$ = 0.151 $\pm$ 0.048 (GeV)$^{2}$ with
$\chi^2$/NDF = 0.71.  The parameter $T$ slowly decreases with the collision energy.

\begin{figure}[htp] 
\center
\includegraphics[width=0.65\linewidth]{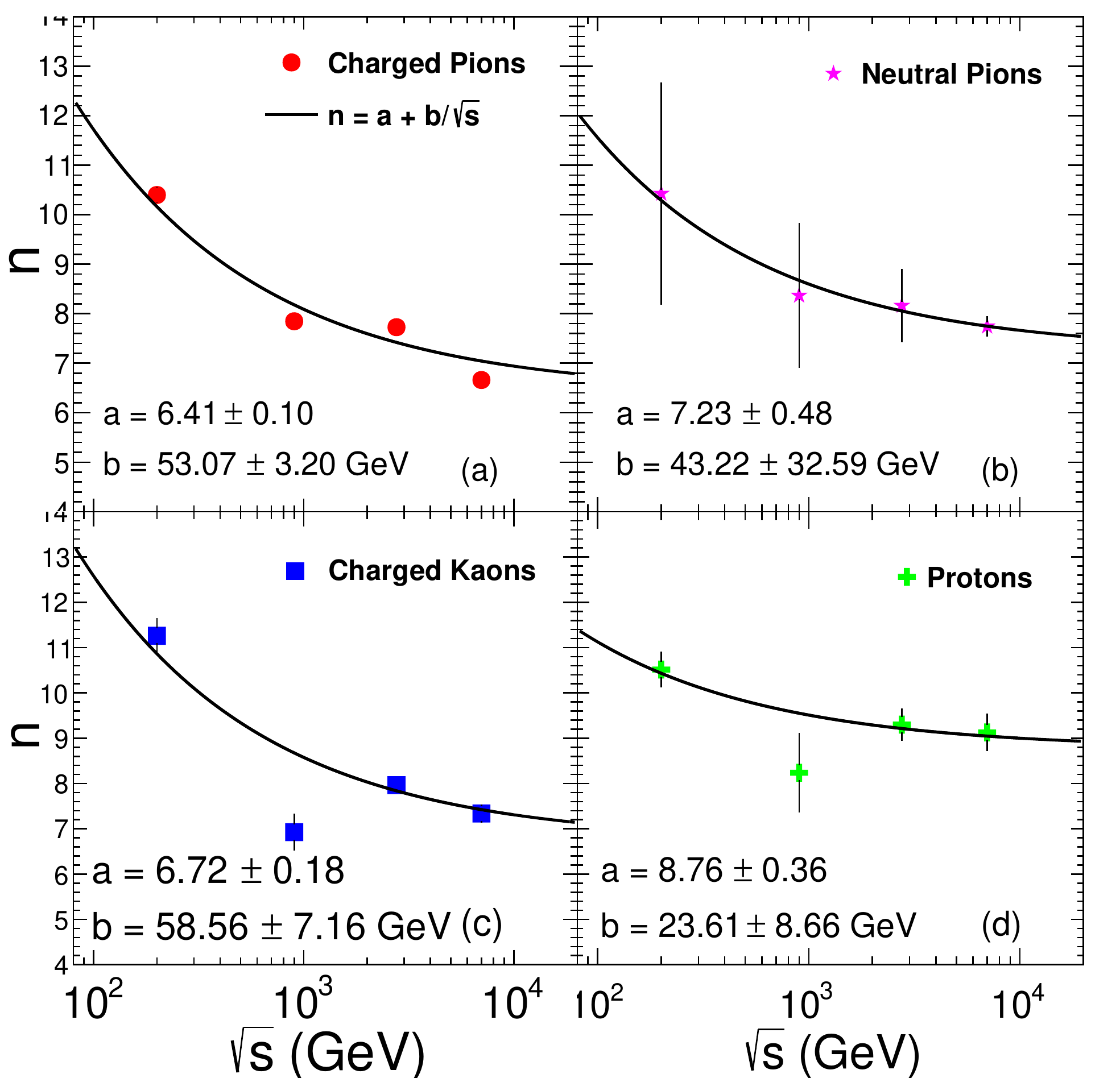}
\caption{ The Tsallis parameter $n$ as a function of the centre of mass energy 
$\sqrt{s}$ of pp collision for (a) charged pions (b) neutral pions (c) charged 
kaons and (d) protons. The solid curve represents the function 
($a + b/\sqrt{s}$).}
\label{Figure6_identifie_pp_collision_nn}
\end{figure}

\begin{figure}[htp] 
\center
\includegraphics[width=0.65\linewidth]{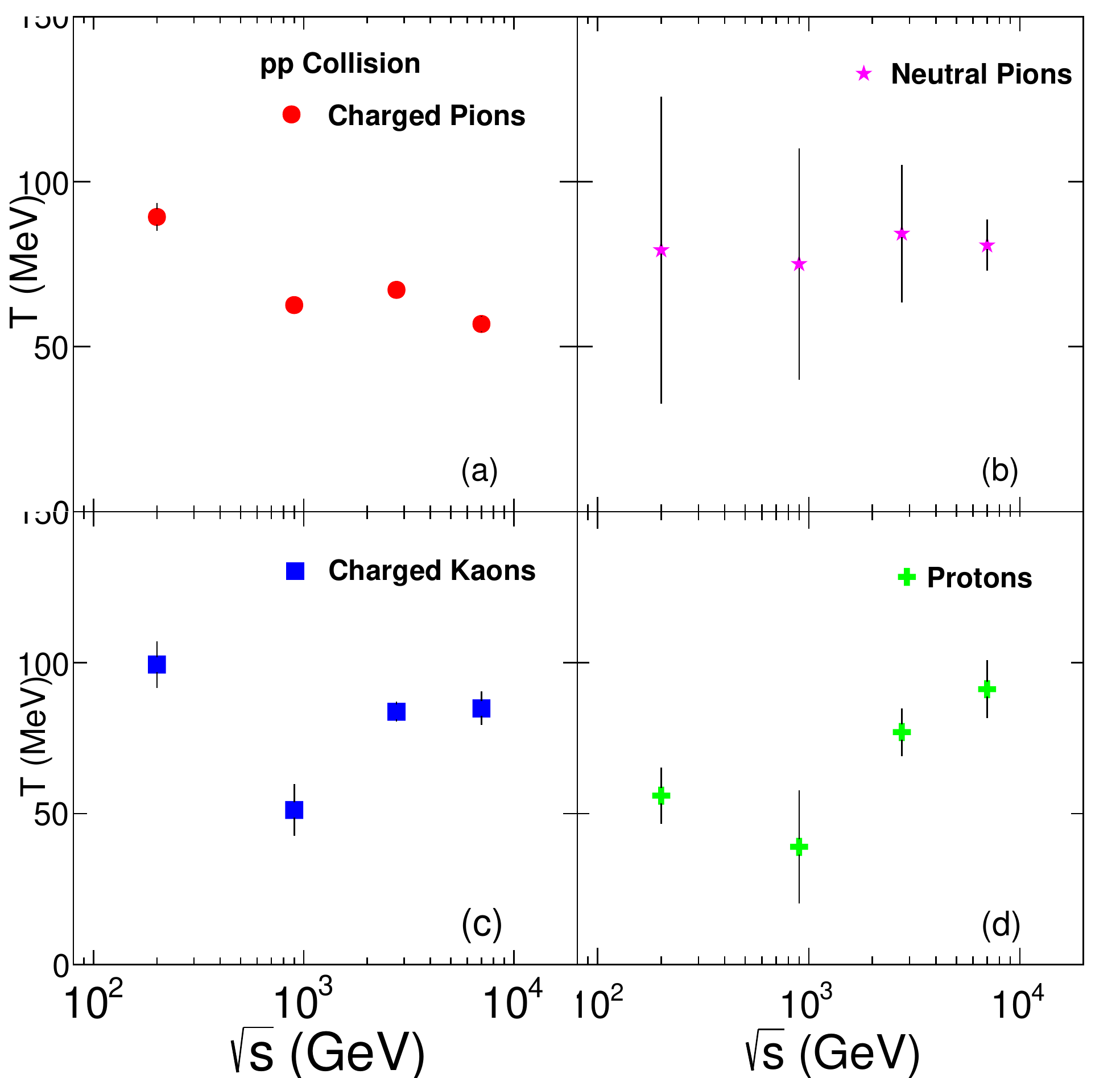}
\caption{ The Tsallis parameter $T$ as a function of the centre of mass energy  
$\sqrt{s}$ of pp collision for (a) charged pions (b) neutral pions (c) charged 
kaons and (d) protons.}
\label{Figure7_identifie_pp_collision_TT}
\end{figure}

 We also present such analysis for each identified particle separately although in a much
 smaller range of $p_{\rm{T}}$ limited by measured data.
 The identified particle $p_{\rm{T}}$ spectra measured at RHIC and LHC energies
\cite{Adare:2011vy, Agakishiev:2011dc, Aamodt:2011zj, Abelev:2014laa, Adam:2015qaa, Adler:2003pb, Abelev:2012cn, Abelev:2014ypa} are fitted with the Tsallis distribution and its parameters $n$ and $T$ are obtained.
 Figure \ref{Figure6_identifie_pp_collision_nn} shows the Tsallis parameter $n$ for the
identified charged particles as a function of $\sqrt{s}$. The
panel (a) is for charged pions, (b) is for neutral pions, (c) is for charged kaons and
(d) is for protons. The solid curve is the fit given by function in
Eq. \ref{nn_parameterized_equation}. The asymptotic values of parameter $n$ are 
6.41 $\pm$ 0.10 for charged pions, 7.23 $\pm$ 0.48 for neutral pions, 6.72 $\pm$ 0.18 for 
charged kaons and 8.76 $\pm$ 0.36 for protons.
The value of $\chi^2$/NDF is not good for charged pions and kaons because the value of $n$
at $\sqrt{s}$ = 900 GeV is away from the fit line.
 This study shows that the asymptotic value of $n$ for unidentified charged particles is
closer to what is obtained for pions and kaons.
Figure \ref{Figure7_identifie_pp_collision_TT}  shows the Tsallis temperature parameter $T$ for
the identified particles as a function of $\sqrt{s}$ in pp collision.
 The panel (a) is for charged pions, (b) is for neutral pions, (c) is for charged kaons and
(d) is for protons.
The bahavior of the parameter $T$ is similar to what is
shown in Fig.~\ref{Figure5_pp_collision_TT}.

\begin{figure}[htp] 
\center
\includegraphics[width=0.65\linewidth]{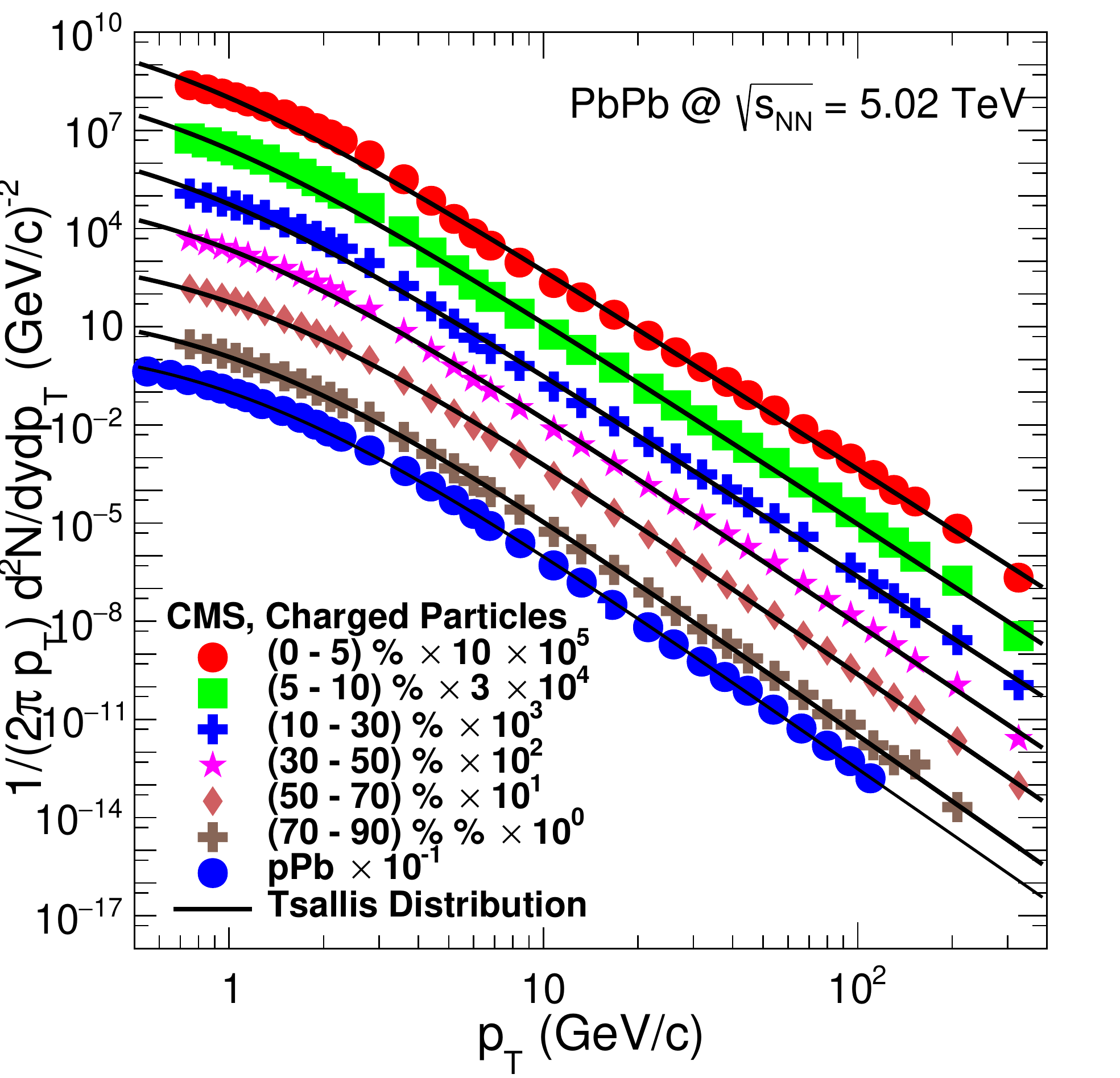}
\caption{The invariant yields of the charged particles  as a function of the transverse  
momentum $p_{\rm{T}}$ for pPb collisions and different centralities of PbPb collisions at 
$\sqrt{s_{\rm{NN}}}$ = 5.02 TeV measured by the CMS \cite{Khachatryan:2016odn, Khachatryan:2015xaa}. 
The solid curves are the fitted Tsallis distribution functions (Eq.~\ref{Tsallis}).}
\label{Figure8_pbpb_502tev_cms}
\end{figure}

\begin{figure}[htp] 
\center
\includegraphics[width=0.65\linewidth]{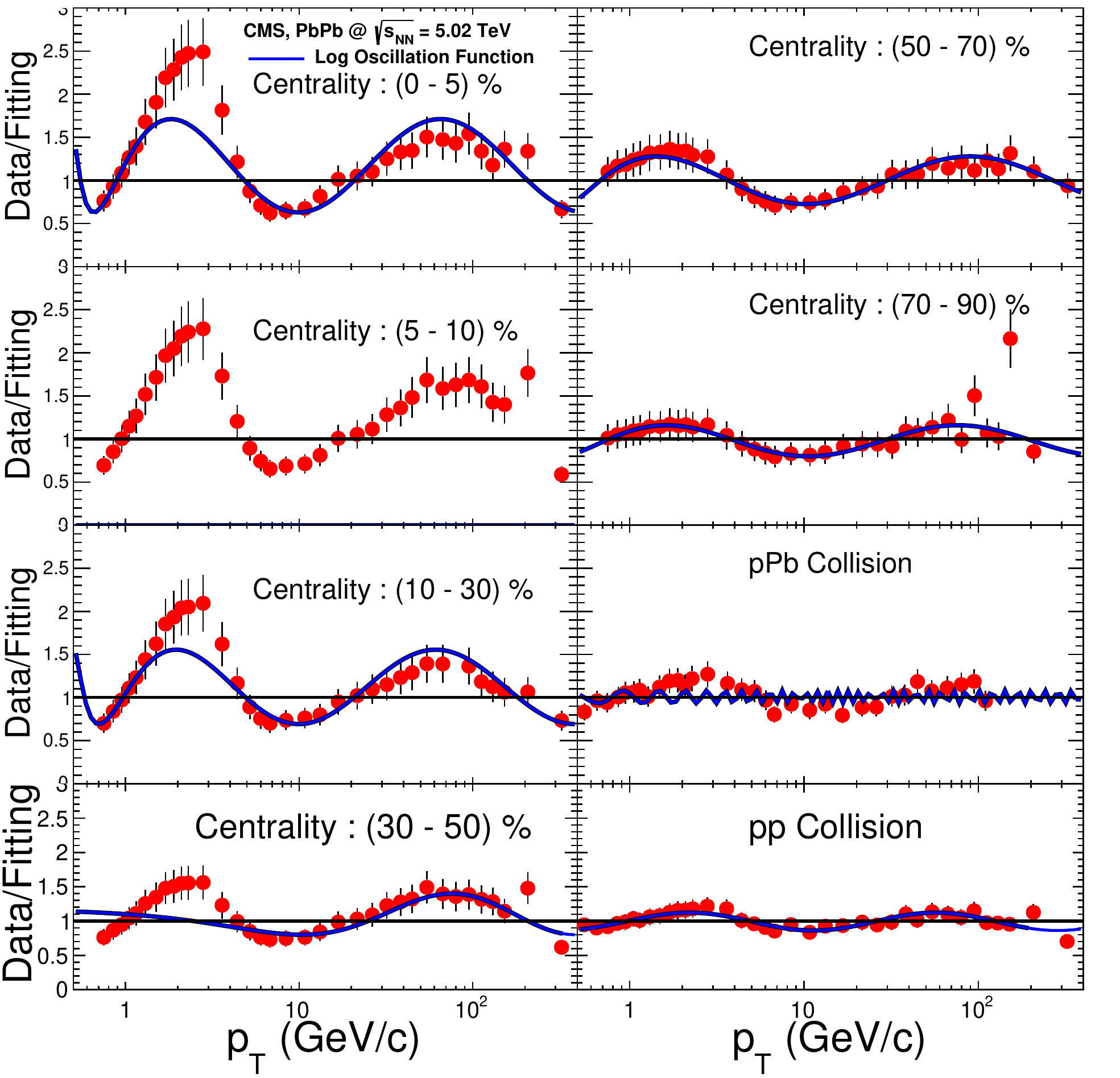}
\caption{The ratio of the charged particle yields data and their Tsallis fits as a function   
of the transverse momentum $p_{\rm{T}}$ for pp, pPb and PbPb collisions at $\sqrt{s_{\rm{NN}}} = $
5.02 TeV. The solid curves are given by Eq.~\ref{log_oscillation_function_equation}.}
\label{Figure9_pbpb_502tev_cms_databyfit}
\end{figure}

  Figure \ref{Figure8_pbpb_502tev_cms} shows the invariant yields of the charged particles
as a function of $p_{\rm{T}}$ for pPb collisions and for many centralities of PbPb collisions
at $\sqrt{s_{\rm{NN}}}$ = 5.02 TeV measured by the CMS experiment \cite{Khachatryan:2016odn}.
The solid curves are the fitted Tsallis distributions (Eq.~\ref{Tsallis}).
 Figure~\ref{Figure9_pbpb_502tev_cms_databyfit}
shows the ratio of the data and the fitted Tsallis distribution as a  function of $p_{\rm{T}}$
for pp, pPb and PbPb collisions at $\sqrt{s_{\rm{NN}}}$ = 5.02 TeV. As we move from peripheral
to central PbPb collisions, the data show increasing deviations from the Tsallis fit.
The $\chi^{2}/\rm{NDF}$ values of the Tsallis fit are given in 
the Table~\ref{pbpb_collision_276tev_chi2_NDF}. 
  The pp and pPb data also show some deviations from the fit.
Interestingly, the deviation pattern in pp and pPb looks similar to that in PbPb collisions
with deviation magnitude increasing with system size. It was suggested in
Ref.~\cite{Wilk:2014bia, Rybczynski:2014ura} that the ratio of the data to the Tsallis fit
shows a log oscillation function and which can be parametrized by a function of the form
\begin{eqnarray}
f(p_{\rm{T}}) = a + b~ \cos\big[c~\log(p_{\rm{T}} + d) + e\Big]~. 
\label{log_oscillation_function_equation}
\end{eqnarray}
Here $a$, $b$, $c$, $d$ and $e$ are the fit parameters. We observe from
Fig.~\ref{Figure9_pbpb_502tev_cms_databyfit} that the log oscillation function although loaded
with five parameters does not describe the deviation pattern specially for more central PbPb
collisions.

\begin{figure}[htp] 
\center
\includegraphics[width=0.65\linewidth]{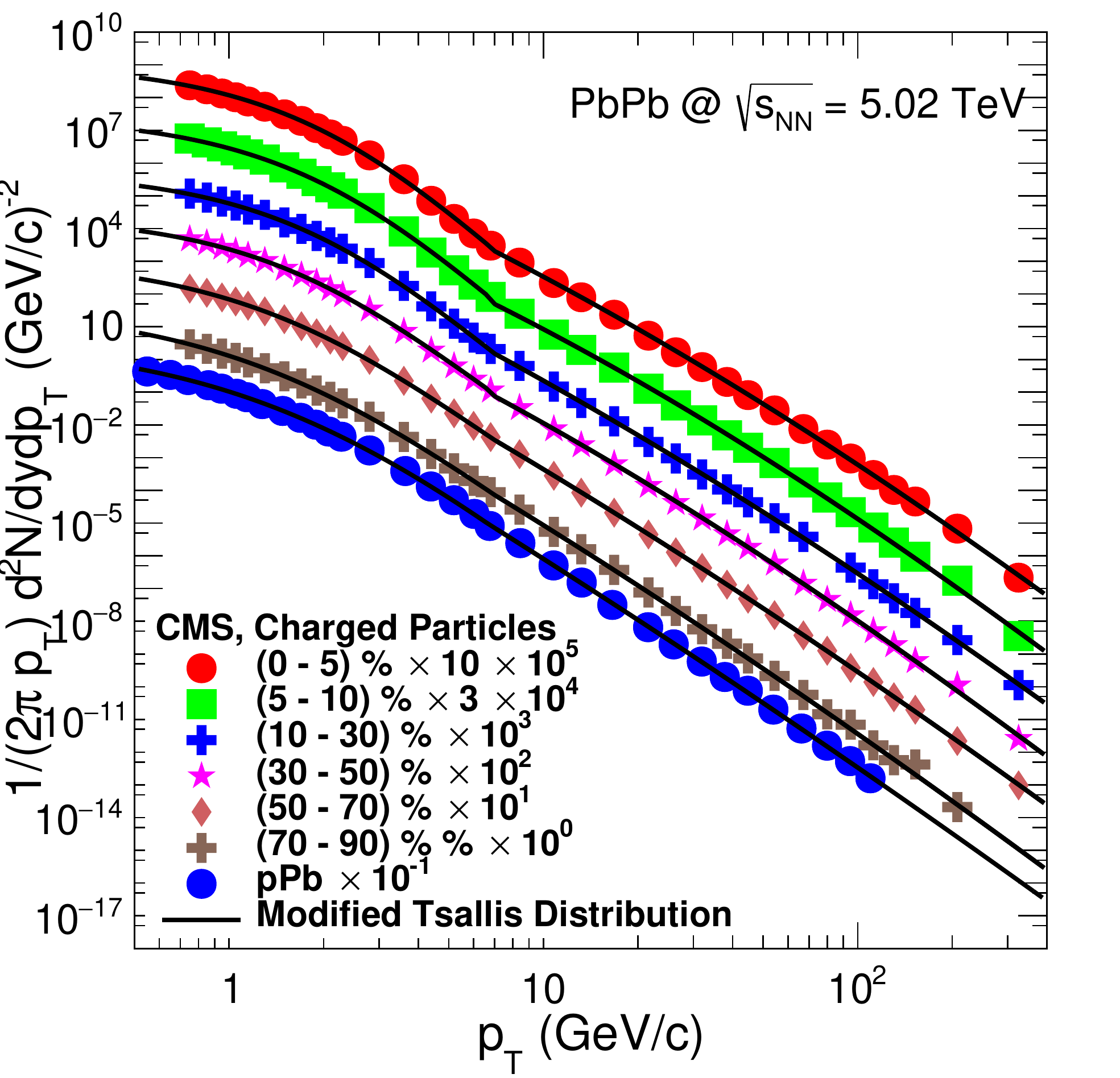}
\caption{The invariant yields of the charged particles  as a function of the transverse 
momentum $p_{\rm{T}}$ for pPb and PbPb collisions at $\sqrt{s_{\rm{NN}}}$ = 5.02 TeV measured by 
the CMS \cite{Khachatryan:2016odn, Khachatryan:2015xaa}. The solid curves are the modified 
Tsallis distributions (Eq. \ref{modified_new_func_tsallis_distribution_function}).}
\label{Figure10_pbpb_502tev_cms_new}
\end{figure}

\begin{figure}[htp] 
\center
\includegraphics[width=0.65\linewidth]{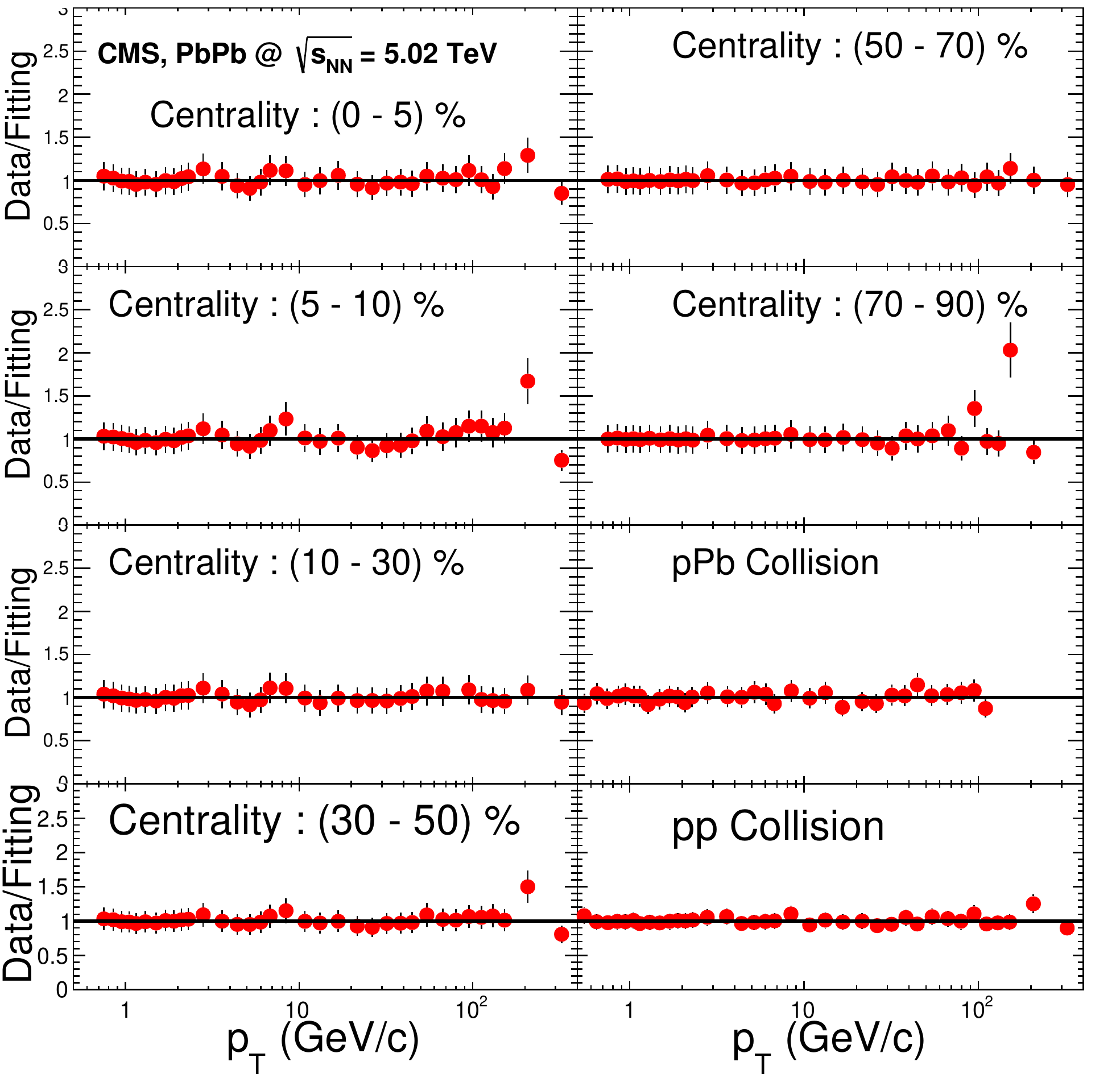}
\caption{The ratio of the charged particle yield data and the fit function (Modified Tsallis  
distribution Eq.~\ref{modified_new_func_tsallis_distribution_function}) as a function of the 
transverse momentum $p_{\rm{T}}$ for pp, pPb and PbPb collisions at $\sqrt{s_{\rm{NN}}}$ = 5.02 TeV.}
\label{Figure11_pbpb_502tev_cms_databyfit_new}
\end{figure}

 Figure \ref{Figure10_pbpb_502tev_cms_new} shows the invariant yields of the charged particles as
a function of $p_{\rm{T}}$ for pPb and PbPb collisions at $\sqrt{s_{\rm{NN}}}$ = 5.02 TeV measured by the CMS
experiment \cite{Khachatryan:2016odn, Khachatryan:2015xaa}. The solid curves are the modified Tsallis
distributions given by Eq.~\ref{modified_new_func_tsallis_distribution_function}.
Figure \ref{Figure11_pbpb_502tev_cms_databyfit_new} shows the ratio of the data and the fit function by
the modified Tsallis distribution as a function
of $p_{\rm{T}}$ for pp, pPb and PbPb collisions at $\sqrt{s_{\rm{NN}}}$ = 5.02 TeV. The ratio of the data and
the fit function shows that the modified Tsallis distribution function gives excellent description of
the measured data in full $p_{\rm{T}}$ range for all the systems. The parameters of the modified Tsallis 
distribution are given in the Table~\ref{table_for_new_tsallis_function}. The values of the first
set of parameters ($n_1$, $p_1$, $\beta$) increase with increasing system size for PbPb collisions.
It shows that degree of thermalization (governed by $n_1$) and the transverse flow (governed by $\beta$)
increase with system size. While fitting the second function, we fix the parameter $n_{2}=7.7$ guided 
by pp value.
  The exponent $\alpha$ which decides the variation of the energy loss of partons as a
function of their energy remains within 0.4 to 0.7. The parameter $B$ is proportional to system
size and increases as we move from pp to the most central PbPb collisions.
To summarize, the functions given in Eqs.~\ref{modified_new_func_tsallis_distribution_function} give
excellent description of the hadron spectra over wide range of $p_{\rm{T}}$ with its parameters indicating
different physics effects in the collisions.

The modified function (Eq.~\ref{modified_new_func_tsallis_distribution_function}) is
intended for PbPb systems to cover medium effects. We have applied it for small
systems as well.
  The charged particle spectrum in pp collision when fitted with the
Tsallis (Eq.~\ref{Tsallis}) gives $\chi^{2}/\rm{NDF}=0.79$.  
 The $\chi^{2}/\rm{NDF}$ value improves to 0.31 when fitting is performed with
 the modified Tsallis (Eq.~\ref{modified_new_func_tsallis_distribution_function}).
 The pp system shows a small transverse flow and energy loss effect.
 It is not surprising since recent experiments have measured effects
 of collectivity in pp collisions when analysis on high muliplicity events was
 performed \cite{Khachatryan:2016txc}.

\begin{table}[ht]
\caption{The parameters of the Tsallis function obtained by fitting the 
  charged particle spectra in pp collision at RHIC and LHC energies.}
\begin{center}
\begin{tabular}{|c || c | c | c | c | c |}  \hline
$\sqrt{s}$   & Experiment  & $n$ & $q$ & $T$  & $\frac{\chi^{2}}{\rm{NDF}}$ \\ 
    &             &                  &                  &       (MeV)        &           \\ \hline \hline
    
62.4 GeV   &  PHENIX   & 14.28 $\pm$ 0.69 & 1.07 $\pm$ 0.05 & 102.87 $\pm$ 5.98 & 0.49 \\ \hline 

 200 GeV   &  PHENIX   & 11.13 $\pm$ 0.45 & 1.09 $\pm$ 0.04 & 98.27 $\pm$ 6.50 & 0.49 \\ \hline 

 900 GeV   &  ALICE   & 8.69 $\pm$ 0.13 & 1.12 $\pm$ 0.02 & 84.71 $\pm$ 2.58 & 0.16 \\ \hline 

 2.76 TeV   &  ALICE   & 7.95 $\pm$ 0.07 & 1.13 $\pm$ 0.01 & 84.57 $\pm$ 2.21 & 0.17 \\ \hline 

 7 TeV   &  ALICE   & 7.48 $\pm$ 0.05 & 1.13 $\pm$ 0.01 & 84.81 $\pm$ 2.08 & 0.26 \\ \hline 

 13 TeV   &  ALICE   & 7.03 $\pm$ 0.14 & 1.14 $\pm$ 0.02 & 81.40 $\pm$ 3.20 & 0.38 \\ \hline 

 900 GeV   &  CMS   & 8.72 $\pm$ 0.22 & 1.11 $\pm$ 0.03 & 83.24 $\pm$ 4.19 & 0.12 \\ \hline 

 2.36 TeV   &  CMS   & 7.79 $\pm$ 0.36 & 1.13 $\pm$ 0.05 & 78.61 $\pm$ 6.82 & 0.21 \\ \hline 

 2.76 TeV   &  CMS   & 7.98 $\pm$ 0.21 & 1.13 $\pm$ 0.03 & 89.66 $\pm$ 13.64 & 0.02 \\ \hline 

 5.02 TeV   &  CMS   & 7.67 $\pm$ 0.02 & 1.13 $\pm$ 0.00 & 87.58 $\pm$ 2.26 & 0.79 \\ \hline 

 7 TeV   &  CMS   & 7.55 $\pm$ 0.16 & 1.13 $\pm$ 0.02 & 86.46 $\pm$ 12.60 & 0.02 \\ \hline 
\end{tabular}
\end{center}
\label{pp_collision_chi2_NDF}
\end{table}

\begin{table}[ht]
\caption{The $\chi^{2}/\rm{NDF}$ of the Tsallis function obtained by fitting the 
  charged particle spectra in pp, pPb and PbPb collisions at $\sqrt{s_{\rm{NN}}}$= 5.02 TeV.}
\begin{center}
\begin{tabular}{|c || c | c | c | c | c | c  | c | c |}  \hline
System                     & PbPb   &  PbPb   & PbPb    &  PbPb   &  PbPb    &  PbPb    & pPb  & pp    \\ \hline 
Centrality ($\%$)          & 0  - 5 & 5  - 10 & 10 - 30 & 30 - 50 &  50 - 70 & 70 - 90  & -    &  -    \\  \hline 
$\frac{\chi^{2}}{\rm{NDF}}$ & 5.93   & 6.11    & 3.89    & 3.11    & 1.84     & 1.20     & 1.22 & 0.79  \\ \hline 
\end{tabular}
\end{center}
\label{pbpb_collision_276tev_chi2_NDF}
\end{table}

\begin{table}[ht]
\caption{The parameters of the modified Tsallis function Eq.~\ref{modified_new_func_tsallis_distribution_function} 
obtained by fitting the charged particle spectra in pp, pPb and PbPb collisions at $\sqrt{s_{\rm{NN}}}$=5.02 TeV. }
\begin{center}
\begin{tabular}{|c || c | c | c | c | c | c  |}  \hline
 System     & $n_{1}$       & $p_{1}$         & $\beta$         &  $\alpha$      & $B$       & $\frac{\chi^{2}}{\rm{NDF}}$  \\  
            &              & (GeV/$c$)       &                 &                & (GeV/$c$)      & \\ \hline \hline
    PbPb       & 8.28 $\pm$ 0.92 & 1.38 $\pm$  0.17 & 0.62 $\pm$ 0.06 &  0.47 $\pm$  0.03 & 5.72 $\pm$ 0.23 & 0.29 \\ 
(0 - 5 $\%$)   &                 &                  &                 &                   &                 &      \\  \hline
   PbPb        & 8.13 $\pm$ 0.93 & 1.35 $\pm$  0.18 & 0.60 $\pm$ 0.06 &  0.42 $\pm$  0.04 & 5.45 $\pm$ 0.33 & 0.68 \\ 
(5 - 10 $\%$)  &                 &                  &                 &                   &                 &      \\  \hline
   PbPb        & 7.73 $\pm$ 0.81 & 1.28 $\pm$ 0.14 & 0.62 $\pm$ 0.08  &  0.52 $\pm$ 0.03  & 5.01 $\pm$ 0.20 & 0.14 \\ 
(10 - 30 $\%$) &                 &                  &                 &                   &                 &      \\  \hline
   PbPb :      & 7.03 $\pm$ 0.65 & 1.11 $\pm$  0.11 & 0.61 $\pm$ 0.11 &  0.45 $\pm$  0.04 & 4.25 $\pm$ 0.22 & 0.36 \\  
(30 - 50 $\%$) &                 &                  &                 &                   &                 &      \\  \hline
   PbPb        & 6.64 $\pm$ 0.72 & 0.96 $\pm$ 0.12  & 0.50 $\pm$ 0.08 &  0.58 $\pm$ 0.03  & 3.73 $\pm$ 0.17 & 0.06 \\ 
(50 - 70 $\%$) &                 &                  &                 &                   &                 &      \\  \hline
   PbPb        & 6.66 $\pm$ 1.20 & 0.90 $\pm$ 0.26  & 0.34 $\pm$ 0.06 &  0.59 $\pm$ 0.04  & 2.95 $\pm$ 0.16 & 0.63 \\  
(70 - 90 $\%$) &                 &                  &                 &                   &                 &      \\  \hline
   pPb         & 7.78 $\pm$ 1.70 & 1.34 $\pm$ 0.95  & 0.14 $\pm$ 0.05 &  0.66 $\pm$ 0.08  & 2.95 $\pm$ 0.73 & 0.32 \\  
               &                 &                  &                 &                   &                 &      \\  \hline
   pp          & 7.78 $\pm$ 1.70 & 1.09 $\pm$ 0.41 & 0.14 $\pm$ 0.07  &  0.60 $\pm$ 0.03  & 2.87 $\pm$ 0.10 & 0.31 \\
               &                 &                  &                 &                   &                 &      \\  \hline
\end{tabular}
\end{center}
\label{table_for_new_tsallis_function}
\end{table}

\clearpage 
\section{Conclusion}
We carried out an analysis of transverse momentum spectra of the unidentified charged
particles in pp collisions at RHIC and LHC energies from $\sqrt{s}$ = 62.4 GeV to 13 TeV
using Tsallis distribution function. The power law of Tsallis/Hagedorn form gives very good 
description of the hadron spectra in the $p_{\rm{T}}$ range 0.2 to 300 GeV/$c$.  The power index
$n$ of the $p_{\rm{T}}$ distributions is found to follow a function of the type $a+b/\sqrt{s}$ with
asymptotic value $a = 6.81$. The Tsallis parameter $T$ governing the soft bulk
contribution to the spectra also decreases slowly with the collision energy. We also
provide a Tsallis fit to the $p_{\rm{T}}$ spectra of hadrons in different centralities of PbPb
collisions at $\sqrt{s_{\rm{NN}}}$ = 5.02 TeV. The measured charged particle $p_{\rm{T}}$ spectra in
PbPb collisions show deviations from Tsallis form which become more pronounced as the system size
increases. We suggest simple modifications of the Tsallis function incorporating transverse flow
in the low $p_{\rm{T}}$ region and in-medium energy loss in the high $p_{\rm{T}}$ region. This function
gives excellent description of charged particle spectra in pp, pPb and PbPb collisions with its
parameters having potentials to quantify various in-medium effects in all systems.

\end{document}